\documentclass[reprint,aps,prb,twocolumn,superscriptaddress]{revtex4}
\usepackage{amssymb}
\usepackage{upgreek}
\usepackage{graphicx}
\usepackage{mathrsfs}
\usepackage{textcomp}
\usepackage[utf8]{inputenc}
\usepackage{color}
\usepackage{siunitx}
\usepackage{url}

\usepackage[utf8]{inputenc}
\usepackage{float}
\usepackage[paperwidth=210mm,paperheight=297mm,centering,hmargin=2cm,vmargin=2cm]{geometry}
\usepackage{lipsum}
\usepackage{circuitikz}
\usepackage{fancyhdr}
\usepackage{mathtools}
\usepackage{bm}
\usepackage[normalem]{ulem}
\usepackage{upgreek}

\begin{document}

\title{Nematicity in the superconducting mixed state of strain detwinned underdoped $\bm{\text{Ba}(\text{Fe}_{1-x}\text{Co}_x)_2\text{As}_2}$}

\author{J. Schmidt}
\affiliation{Universidad de Buenos Aires, FCEyN, Departamento de F{\'\i}sica. Buenos Aires 1428, Argentina.}
\affiliation{CONICET-Universidad de Buenos Aires, IFIBA, Buenos Aires 1428, Argentina.}
\author{V. Bekeris}\affiliation{Universidad de Buenos Aires, FCEyN, Departamento de F{\'\i}sica. Buenos Aires 1428, Argentina.}
\affiliation{CONICET-Universidad de Buenos Aires, IFIBA, Buenos Aires 1428, Argentina.}

\author{G. S. Lozano}
\affiliation{Universidad de Buenos Aires, FCEyN, Departamento de F{\'\i}sica. Buenos Aires 1428, Argentina.}
\affiliation{CONICET-Universidad de Buenos Aires, IFIBA, Buenos Aires 1428, Argentina.}
\author{M. V. Bortulé}
\affiliation{Universidad de Buenos Aires, FCEyN, Departamento de F{\'\i}sica. Buenos Aires 1428, Argentina.}
\author{M. Marziali Bermúdez}
\affiliation{Universidad de Buenos Aires, FCEyN, Departamento de F{\'\i}sica. Buenos Aires 1428, Argentina.}
\affiliation{CONICET-Universidad de La Plata, IFLYSIB, La Plata 1900, Argentina}
\author{C. W. Hicks}
\affiliation{Max Planck Institute for Chemical Physics of Solids, Nöthnitzer Str 40, 01187 Dresden, Germany}
\author{P. C. Canfield}
\affiliation{Ames Laboratory and Department of Physics and Astronomy, Iowa State 
University, Ames, Iowa 50011, USA}
\author{E. Fradkin}
\affiliation{ Department of Physics and Institute for Condensed Matter Theory, University of Illinois at Urbana-Champaign, 1110 West Green Street, Urbana, Illinois, 61801-3080, USA} 
\author{G. Pasquini}
\affiliation{Universidad de Buenos Aires, FCEyN, Departamento de F{\'\i}sica. Buenos Aires 1428, Argentina.}
\affiliation{CONICET-Universidad de Buenos Aires, IFIBA, Buenos Aires 1428, Argentina.}

\date{\today }

\pacs{1234}

\begin{abstract}

Evidence of nematic effects in the mixed superconducting phase of slightly underdoped $\text{Ba}(\text{Fe}_{1-x}\text{Co}_x)_2\text{As}_2$ is reported. We have found strong in-plane resistivity anisotropy for crystals in different strain conditions. For these compositions, there is no magnetic long range order, so the description may be ascribed to the interplay between the superconducting and nematic order parameters. A piezoelectric-based apparatus is used to apply tensile or compressive strain to tune nematic domain orientation in order to examine intrinsic nematicity. Measurements are done under a rotating magnetic field and the analysis of the angular dependence of physical quantities identifies the cases in which the sample is {\em detwinned}. Furthermore, the angular dependence of the data allows us to evaluate the effects of nematicity on the in-plane superconductor stiffness. Our results show that although nematicity contributes in a decisive way to the conduction properties, its contributions to the anisotropy properties of the stiffness of the superconducting order parameter is not as significant in these samples.

\end{abstract}

\maketitle

\section{Introduction} 
\label{sec:Introduction}

The role of electronic nematicity \cite{Kivelson98, Fradkin99} in unconventional
superconductivity has been theoretically explored in terms of coupling
between the nematic and superconducting order parameters \cite{Fradkin2010, Fradkin2015}. Growing experimental results pointing towards this connection have been published during the last decade. In particular, an anisotropic phase has been reported in the underdoped regime of both cuprate \cite{Cupratos} and Fe-based \cite{otrosFBS, Tanatar2016, Chu2010, Tanatar2010} high-temperature superconductors with a concurrent breaking of the $C_4$ symmetry in the structural and transport properties. 
 
 Recent Raman \cite{Raman} and elastoresistivity \cite{Chu2012, Kuo2013, Kuo2016} experiments in the $\text{Ba}(\text{Fe}_{1-x}\text{Co}_x)_2\text{As}_2$ family have established that this simultaneous symmetry breaking is driven by electronic degrees of freedom, consistent with the existence of a true nematic phase. In many superconducting compounds this phase develops until the system undergoes a transition to an antiferromagnetic order at a lower temperature. In underdoped pnictides, the most conclusive experimental observation supporting the interplay between nematic and superconducting order is the fact that their phase boundaries intersect at a composition $x_\text{c}$ near the optimal doping $x_\text{op}$ \cite{Ni2008, Prosorov2009}, where the signature of a nematic quantum critical point has been reported \cite{Kuo2016}. In the $\text{Ba}(\text{Fe}_{1-x}\text{Co}_x)_2\text{As}_2$ family,  $x_\text{c}\approx 0.067 \lesssim  x_\text{op}\approx 0.074$.

Recently, a \textit{nematic superconducting phase} in an optimally doped tetragonal compound of the family $\text{Ba}_{1-x}\text{K}_x\text{Fe}_2\text{As}_2$ \cite{Li2018} was reported. The strong symmetry breaking in the superconducting transport properties, in contrast to the very weak symmetry breaking in the normal phase may originate from the strong quantum fluctuations of a nearby nematic quantum critical point, as found in recent state-of-the-art quantum Monte Carlo simulations.\cite{lederer-2017}

On the other hand, reentrant magnetic \cite{Fernandes2010} and orthorhombic-tetragonal \cite{Nandi2010} transitions have been reported in the superconducting phase of underdoped $\text{Ba}(\text{Fe}_{1-x}\text{Co}_x)_2\text{As}_2$, the latter occuring near optimal doping. These facts, together with an enhancement of the superfluid density in nematic domain boundaries (DBs) \cite{Kalisky2010}, in agreement with a repulsion of superconducting vortices \cite{Kalisky2011}, are all evidence suggesting a  competition between nematicity and superconductivity in these compounds.

In fact, slightly underdoped $\text{Ba}(\text{Fe}_{1-x}\text{Co}_x)_2\text{As}_2$ single crystals \cite{canfield2010} are ideal compounds to study the interplay between nematic and superconducting order parameters. For these doping concentrations, the spontaneous orthorhombic distortion is very small (less than $0.05\%$ for $x = 0.062$) \cite{Nandi2010}, whereas the elastoresistivity near the structural transition is huge \cite{Kuo2016}. Moreover, the superconducting transition occurs in the
absence of any competing magnetic order \cite{Ni2008}. Therefore, the symmetry of the superconducting properties near the critical temperature can bring valuable information on the possible coupling with the nematic phase. 

In this framework, the formation of a dense array of nematic domains at submicron distances in typical as-grown crystals, is a crucial issue. It has been shown that in slightly underdoped compounds, with low orthorhombic distortions, the twinning distances are also very short, making domains not observable with standard optical methods \cite{Prosorov2009}. 
Furthermore, the very small structural distortion is 
even undetectable with standard x-ray characterization. However,
the presence of DBs is expected to dramatically modify
vortex physics \cite{Nelson1992, Blatter}, providing a means to recognize the presence of nematic domains without the need for additional experimental techniques. 

The interplay between superconducting vortices and DBs (also associated with structural twin boundaries) has been extensively
investigated in the past in $\text{YBa}_2\text{Cu}_3\text{O}_{7-\delta}$ single crystals \cite{Kwok96, Zhukov97, Grigera98, Herbsommer2000, Pasquini2007} and more recently experiments were performed in underdoped Fe-based compounds \cite{Kalisky2011, Marziali2013, Yagil2016}. DBs indeed can act as a source of correlated disorder, so their presence modifies the superconducting anisotropy expected in a detwinned single domain. Taking advantage of this fact, in the present work, we investigate the symmetry in the superconducting transport properties in a slightly underdoped 
$\text{Ba}(\text{Fe}_{1-x}\text{Co}_{x})_{2}\text{As}_{2}$ single crystal under different strain conditions. 

The effects of anisotropy in superconductors with uniaxial symmetry were analyzed in 1990s within the Ginzburg-Landau formalism, using simple scaling laws that predict the behavior of physical quantities as the relative orientation of the external magnetic field and the $c$ axis is varied \cite{Blatter, BlatterB}. Failure of this scaling is interpreted as the prevalence of correlated disorder, such as that originating from the presence of DBs \cite{Herbsommer2000, Marziali2013, Jaros2008}.

In this work, we measure the transport properties in the superconducting mixed phase, under the out-of-plane rotation of an applied magnetic field. From these measurements, we are able to detect the presence of DBs in free samples as well as the detwinning under the application of strong compressive or tensile strains.  By generalizing the scaling formalism to nematic systems, we determine the in-plane anisotropies of the superconducting stiffness and of transport properties in the mixed phase. We observe a strong in-plane resistivity anisotropy, suggesting that nematicity strongly affects transport properties related to vortex dissipative dynamics. On the other hand, the superconducting stiffness, associated with the energy cost of local changes of the superconducting order parameter, seems to be unaffected by the strain within our experimental resolution.

This paper is organized as follows: in Sec. II we describe the experimental array; results and discussion are presented in Sec. III, and conclusions are drawn in Sec. IV.

\section{Experimental}
\label{sec:Experimental}

Samples used in this work are single
crystals of $\text{Ba}(\text{Fe}_{1-x}\text{Co}_{x})_{2}\text{As}_{2}$,  
grown from FeAs flux from a starting load of metallic Ba, FeAs, and
CoAs, as described in detail elsewhere \cite{Ni2008}. We selected samples with $x = 0.062$, which is very close to optimal doping, and near the maximum doping at which orthorhombicity is observed \cite{Ni2008, Prosorov2009}.
For this Co concentration, the tetragonal-to-orthorhombic phase transition at $T_{s}\simeq 30~\text{K}$ is above the superconducting transition at $T_{c}\simeq 24~\text{K}$, which nucleates in the orthorhombic paramagnetic normal phase with no long-range antiferromagnetic order.
 
\begin{figure}[tb]
 \centering
 \includegraphics[width=\linewidth]{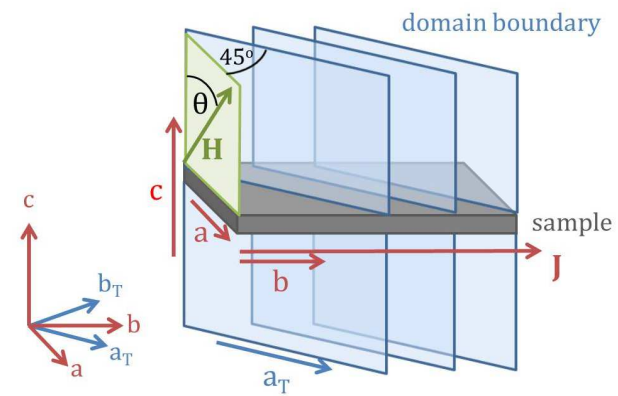}
 \caption{\footnotesize{(Color online) Scheme of the direction of the applied magnetic
field and current. For clarity, just one family of parallel domain boundaries (DBs) is shown. The indicated $a$ and $b$ axes correspond to the first
domain, where the short ${b}$ axis is oriented parallel to the applied
current $\mathbf{J}$. In this domain, the applied magnetic field $\mathbf{H}$ is rotated in the $ac$ plane, forming an angle $\protect\theta$ with the ${
c}$ axis. Current is applied in the $b$ direction. Inset:
Tetragonal ${a_{T}},{b_{T}}$ (blue) and orthorhombic ${a},{b}$ (red) axes are shown (see text).}}
 \label{fig:Direcciones}
\end{figure}

The orientation of the crystalline axes was identified with x-ray
diffractometry in single crystalline platelet samples. Crystals were further
cut along the tetragonal [110] direction into rectangles with a precision
wire saw, so that the $a/b$ orthorhombic axes were parallel to the sample sides upon cooling through the structural transition. In this way,  we expect to be able to detwin the sample through application of compressive or tensile strain along its length. Uniaxial stress can be applied, favoring one orthorhombic orientation over the other; if uniaxial tensile
stress is applied, the longest orthorhombic $a$ axis is favored in
the direction of the applied stress, whereas the shorter $b$ axis
is favored for the case of compressive stress \cite{Chu2010}; for low strains, DBs are expected to form at a $45^{\circ}$ angle, as shown in Fig. \ref{fig:Direcciones}.
 
After cleaving the crystals, current and voltage Au wires  were attached with silver paint on top of Au sputtered contacts along the longest side of the sample. The main results presented in this work correspond to a sample with dimensions of $1.1\times 0.3\times 0.05~\text{mm}^3$. Measurements performed in other samples of the same batch and composition are consistent.
 
Angle-dependent low-current ac magnetotransport experiments were carried out in a Janis continuous flow $^4$He cryostat down to $\simeq 20~\text{K}$ with millikelvin temperature control in the low temperature range, using a lock-in amplifier for audio range frequencies.
 
A dc magnetic field, provided by an electromagnet that can be rotated with a precision of $\simeq 0.5^{\circ }$, was applied perpendicular to the applied current and oriented at a given angle with respect to the sample's ${c}$ axis, as sketched in Fig. \ref{fig:Direcciones}. In Fig. 1 we show an orthorhombic sample with a family of parallel DBs, and the left domain has its shorter ${b}$ axis parallel to the applied current. For clarity, just one family is shown (in real samples many families are present, oriented along both directions, forming a $45^{\circ}$ angle with the sample side). 

 \begin{figure}[tb]
 \includegraphics[width=\linewidth]{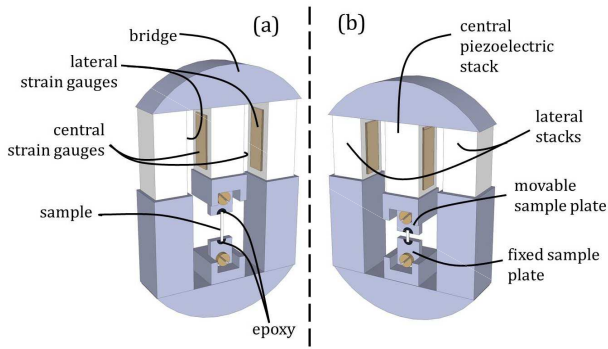}
 \caption{\footnotesize{
 (Color online) (a) and (b) Different components of the apparatus, indicating sample elongation and compression (see text).}}
 \label{fig:Dispositivo}
\end{figure}

In order to control the longitudinal strain, samples were mounted in a recently designed apparatus to tune in-plane uniaxial stress \cite{Hicks2014}. As described in detail in 
{Ref. \citenum{Hicks2014}}, the sample is placed across a gap between two plates joined by a bridge, one of them movable and the other fixed, as sketched in Fig. \ref{fig:Dispositivo}. Three lead zirconium titanate piezoelectric stacks control the position of the movable plate. Compressive or tensile strain is applied by controlling the length differential between the inner and outer stacks, as sketched in Figs. \ref{fig:Dispositivo}(a) and \ref{fig:Dispositivo}(b): a positive voltage applied on the outer two stacks pushes the bridge and elongates the sample, and the opposite occurs if the positive voltage is applied to the central stack. Because the stacks are much longer than the sample, larger strains can be achieved on the sample, compared to other piezoelectric-based straining setups. To measure the displacement applied to the sample, there are strain gauges affixed to the stacks, and these are connected to two opposite branches of a Wheatstone bridge, so that the out-of-balance signal can be measured to determine the gap (or sample) length variation with high resolution. Samples were prepared with high length-to-width and length-to-thickness aspect ratios to increase strain homogeneity, reduce bending, and avoid edge effects \cite{Hicks2014}. Recommendations in the use of epoxy to mount samples and to consider elastic deformation of the mounting epoxy described in Ref \citenum{Hicks2014} were followed. Due to technical restrictions, samples were glued in the asymmetric configuration \cite{Hicks2014}. Any local stress produced by the mounting itself decays at a distance $\lambda \simeq 90~\upmu \text{m}$ and becomes negligible close to the voltage contacts.

Besides the capability to apply larger deformations, this array helps compensate the thermal expansion of the piezoelectric stacks: as all the stacks have equal lengths, similar temperature expansion (contraction) are expected for the inner and outer stacks, so the gap (sample) length should not significantly vary with temperature. However, this thermal compensation is, in practice, not perfect, and in addition, the sample will be strained by differential thermal contraction between it and the frame material (titanium) of the stress cell. The additional strain would need to be compensated by an appropriate voltage on the stacks to achieve the zero strain condition. Notice that most previous experimental works just report a \textit{relative} sample deformation. In this work we measure the {\textit{absolute}} sample strain by applying the
following strategy to determine the zero-strain sample condition: single crystals
are first placed on top of the gap and carefully attached to the sample plates at one end, leaving the other end of the sample free so that it remains unstressed throughout the whole measured temperature. 
We define this arrangement as a \textit{free standing} (F) sample with its
corresponding resistivity $\rho _{\text{F}}(T,H,\theta )$. The next step is to
properly attach the second end of the sample to apply strain, as both ends are driven. We then adjust the voltage applied on the piezo stacks to compensate the apparatus thermal contraction/expansion at each temperature. We consider the sample to have reached the \textit{strain-free} (SF) condition $\epsilon_{\text{SF}} = \epsilon_{\text{F}} = 0$ when the corresponding resistivity $\rho _{\text{SF}}(T)=\rho _{\text{F}}(T)$.
In the rest of the work the longitudinal strain $\epsilon$ refers, in all the cases, to the absolute strain, considering $\epsilon _{\text{SF}}\sim \epsilon_{\text{F}}=0$.

\section{Results and discussion}
\label{sec:Results}

As reported in several works \cite{Chu2010, Tanatar2010, Blomberg2012}, the application of strain in the tetragonal $[1 1 0]$ direction in underdoped $\text{Ba}(\text{Fe}_{1-x}\text{Co}_x)_2\text{As}_2$, promotes a
preferential orientation of nematic domains below $T_s$; high-resolution x-ray studies show that the relative volume fraction of domains with different orientations is bolstered by the applied strain. On the other hand, the applied strain induces in-plane resistivity anisotropy below and above $T_{s}$ due to large elastoresistivity effects. 

\begin{figure}[tbp]
\centering
\includegraphics[width=\linewidth]{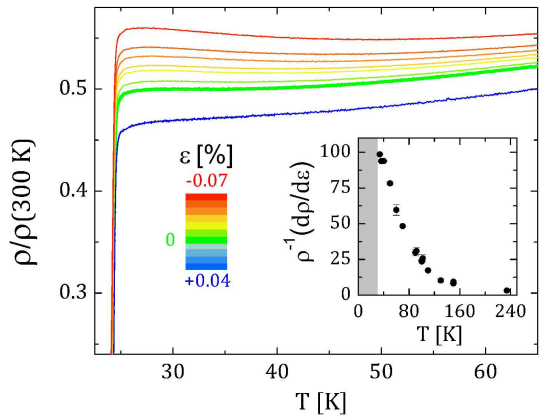}
\caption{\footnotesize{(Color online) Normalized temperature-dependent resistivity of a $%
\text{Ba}(\text{Fe}_{1-x}\text{Co}_{x})_{2}\text{As}_{2}$ single crystal with $x=0.062$ measured for different strain values under uniaxial applied stress in the $[110]$ direction $\protect\varepsilon $. The color code indicates the
strain interval $-0.074<\protect\varepsilon <0.040$. Inset: $\rho^{-1} ~ {d} \rho / {d} \epsilon$ as a function of temperature
in the tetragonal phase, consistent the nematic divergence (see text). }}
\label{fig:Elastores}
\end{figure}

Fig. \ref{fig:Elastores} shows the temperature-dependent resistivity $\rho(T)$ of a $\text{Ba}(\text{Fe}_{1-x}\text{Co}_x)_2\text{As}_2$ sample with $x=0.062$ at temperatures below $70\ \text{K}$, normalized to its value at $T = 300~\text{K}$, measured under different values of applied uniaxial strain $\varepsilon$ in the interval $ -0.07 \leq \varepsilon \leq 0.04$. Negative strain stands for sample compression, while positive strain values indicate tensile strain. For this doping level, the fingerprint in resistivity related to the tetragonal-to-orthorhombic phase transition \cite{Chu2010} at $T_{s}\simeq 30\ \text{K}$ is very weak because the orthorhombicity is low \cite{Nandi2010}. However, a huge dependence on the applied strain is observed in this temperature range. The inset shows the corresponding elastoresistivity component $\frac{1}{\rho} \frac{{d} \rho}{{d} \epsilon}$ as a function of temperature in the tetragonal phase, 
consistent with the nematic divergence reported in a large number of  Fe-based superconductors and recently related to the ubiquitous signatures of nematic quantum criticality in optimally doped Fe-based superconductors \cite{Kuo2016}. In the context of the present work, this huge elastoresistivity allows for using the resistivity as a tool to identify the zero-strain condition with reasonable resolution. 

\begin{figure}[h]
\centering
\includegraphics[width=\linewidth]{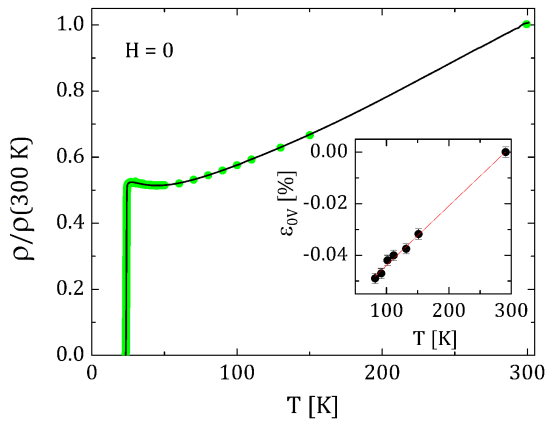}
\caption{\footnotesize{(Color online) Normalized temperature-dependent resistivity of the
freestanding sample (black curve) for $H=0$. The matched resistivity
for the strain-free sample is plotted by green dots. Inset: Sample strain with $0\ V$ on the piezo stacks $\protect\varepsilon _{0\text{V}}$ due to residual differential thermal contraction between the sample and apparatus. The opposite $T$-dependent
strain was applied to the sample to recover resistivity values displayed in black line in
the main panel ( See text).}}
\label{fig:F_SF}
\end{figure}

The procedure described in Sec. \ref{sec:Experimental} was followed to obtain a controlled SF sample as shown in Fig. \ref{fig:F_SF} for another sample at $H=0$. The black cooling curve is the F sample resistivity, measured while the sample had only one of its ends attached to the apparatus. With green dots we plot the matched temperature-dependent resistivity after properly attaching the second end of the sample to the second piezo stack so that both ends were driven independently.  In our particular setup, for zero voltage on the stacks the temperature-dependent resistivity was higher than for the F sample in the whole $T$ range, indicating sample compression.
This compression was quantified and is plotted in the inset.  To recover the free sample resistivity, the opposite strain was applied to the sample.

\begin{figure}[tbp]
\centering
\includegraphics[width=\linewidth]{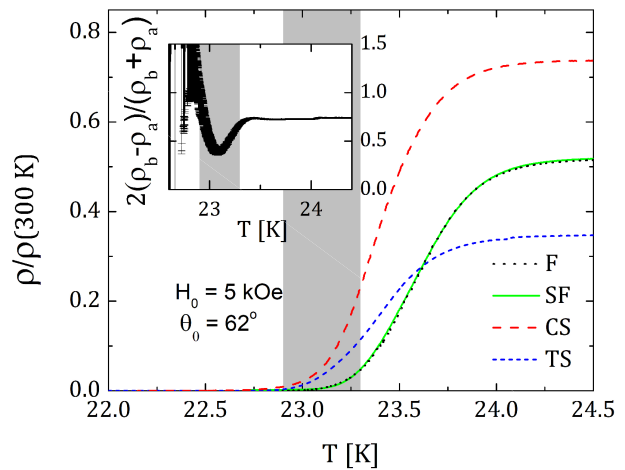} 
\caption{\footnotesize{(Color online) Normalized resistivity transition measured at $H_0 = 5\ \text{kOe}$ applied at $\theta_0 = 62^{\circ}$. The responses of the free standing sample (F, black line) and strain-free sample (SF, green line) are compared with the resistivity measured under a strong tensile strain (TS) $\epsilon_{\text{TS}} = 0.26 \%$ (blue line) and under a strong compressive strain (CS) $\epsilon_{\text{CS}}=-0.35 \% $ (red line). The resistivities measured in TS and CS conditions can be identified as $\rho_a$ and $\rho_b$,  respectively (see text). The inset shows the temperature-dependent resistivity anisotropy, $2(\rho_b -\rho_a)/(\rho_b + \rho_a) > 0.4 $, across the whole resistive transition. The gray regions in both panels identify the temperature range where the effective fields plotted in Fig. \ref{fig:Scaling} were obtained.} }
\label{fig:Trancisiones_strain}
\end{figure}

Central to this work is the connection between electronic nematicity and
superconductivity, so from this point we focus on the resistive superconducting transition. A finite width in the resistivity transition at $H=0$ is expected due to disorder and geometrical effects. Moreover, in an applied magnetic field, there is an additional {transition broadening} due to the
presence of a dissipative vortex liquid in the superconducting phase. 
The main panel of Fig. \ref{fig:Trancisiones_strain}
compares the resistivity transition in a field $H_{0}=5\ \text{kOe}$ applied at a
fixed direction $\theta_{0}$ relative to the ${c}$ axis under different
strain conditions, F and SF procedures (black and green curves), and under strong compressive (CS, red curve) and tensile (TS, blue curve) strains of $-0.35\ \%$ and $0.26\ \%$, respectively. For this comparison, keeping in mind that DBs may be modified by strain,  the magnetic field was applied in a direction  far enough from the $c$ axis and the $a-b$ planes to avoid  vortex pinning by correlated defects. We arbitrarily fixed $\theta_0 = 62 ^{\circ}$, so that random point defects dominate pinning in all the strain conditions. A strain-dependent resistivity is observed along the transition, together with a small decrease in the transition temperature (very small in this case due to
the quasi-optimal sample doping \cite{Budko2009}). 

\begin{figure*}
\centering
\includegraphics[width=0.8\linewidth]{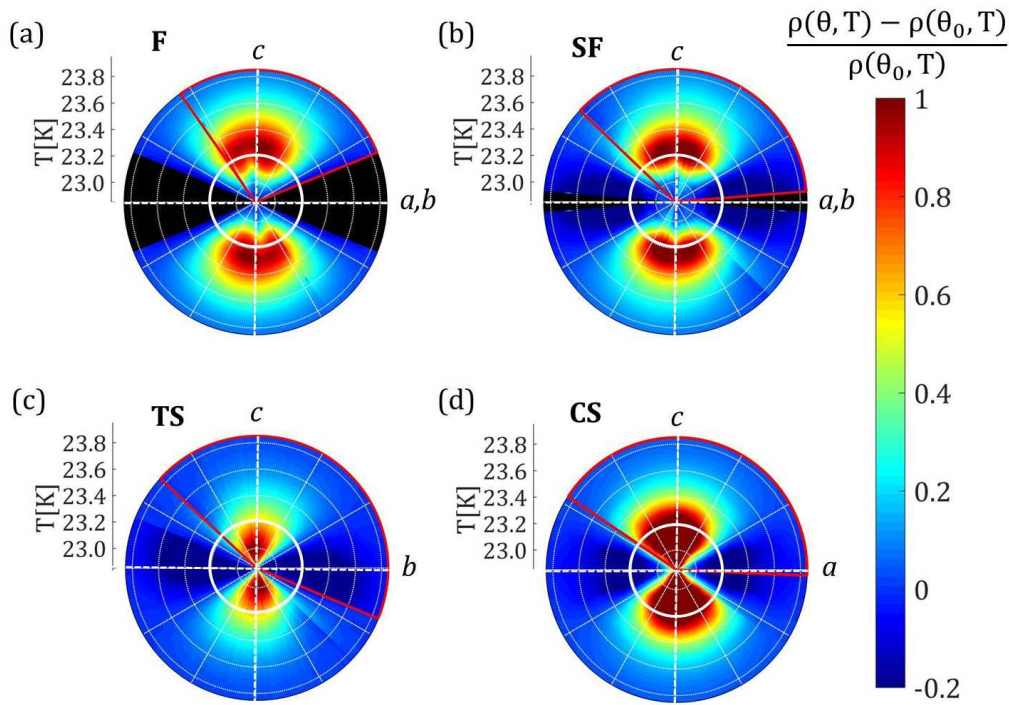}
\caption{\footnotesize{(Color online) Color maps in polar plots for temperature- and angle-dependent normalized magnetoresistivity $[\protect\rho (\protect\theta ,T)-\protect\rho (\protect\theta _{0},T)]/\protect\rho (\protect\theta _{0},T)$ in the superconducting state at $H_{0}=5~\text{kOe}$. Temperature is the radial coordinate,
the angular coordinate is the angle between the applied field and the ${c}$
axis, and the color code indicates the normalized resistivity scale. Results
for different sample strain conditions are shown. (a) Freestanding sample (F). (b)
Strain-free sample (SF). (c) Under tensile strain (TS). (d) Under compressive
strain (CS). The circular segments
delimited by red lines indicate the angular and temperature intervals for
which resistivity was measured, and the plots were completed by symmetry. At low temperatures, delimited by the white circles, an anomaly
is observed for F and SF conditions but absent in strained TS and CS samples.
This anomaly is related to DBs (see text).}}
\label{fig:Polares4}
\end{figure*}

In anisotropic superconductors, a dependence on the magnetic field
direction is expected from the contribution of the vortex liquid
magnetoresistivity $\rho (T,H,\theta )$. This angular dependence gives
therefore information about the underlying anisotropy in the superconducting
phase. Figure \ref{fig:Polares4} summarizes the main experimental results of
this work: polar plots for the angle-dependent normalized magnetoresistivity 
$[\rho (\theta,T )-\rho (\theta _{0},T)]/\rho (\theta _{0},T)$ across the superconducting transition for the four different
sample conditions, F [Fig. \ref{fig:Polares4}(a)], SF [Fig. [\ref{fig:Polares4}(b)], under tensile strain TS [Fig. \ref{fig:Polares4}(c)] and under
compressive strain CS [Fig. \ref{fig:Polares4}(d)]. The color bar represents the normalized
magnetoresistivity, the radial coordinate is the temperature, and the angular
coordinate is the angle between the applied field and the $%
\mathbf{c}$ axis as the field is rotated within a plane perpendicular to the direction of the applied current (see Fig. \ref%
{fig:Direcciones}). The circle sectors delimited by red lines indicate the
angular and temperature intervals for which resistivity was measured.
The plots were completed by symmetry.

The resistivity in the normal state is angle independent within resolution, and
 the angular dependence develops once the dissipation due to the driven
superconducting vortex flow starts playing a role. As expected in a $C_2$ symmetry, in the
interval $[0,\pi /2]$, there is a monotonic decrease of the temperature  where superconductivity nucleates [i.e., the temperature for which $H_{0}=H_{c2}(T,\theta )$] with increasing $\theta $. As the underlying anisotropy is conserved along the transition, a similar monotonic angular dependence is expected for the vortex-liquid-vortex-glass transition at $T_{g}(\theta)$, where the resistivity
drops to zero. However, as can be observed, this is not the case for all the sample conditions; the main feature to be pointed out is the low-resistivity anomaly observed at low temperatures and low angles (encircled in white) in Figs. \ref{fig:Polares4}(a) and \ref{fig:Polares4}(b) that is absent in the strained samples in Figs. \ref{fig:Polares4}(c) and \ref{fig:Polares4}(d). This anomaly is clearly observed in the S and SF phase samples. We attribute the existence of such an anomaly to the presence of planar defects (parallel to the $c$ axis), which act as a source of correlated disorder and strongly influence vortex dynamics \cite{Marziali2013}. The disappearance of this anomaly under the application of stress thus signals the detwinning of the sample.

\begin{figure}
\centering
\includegraphics[width=\linewidth]{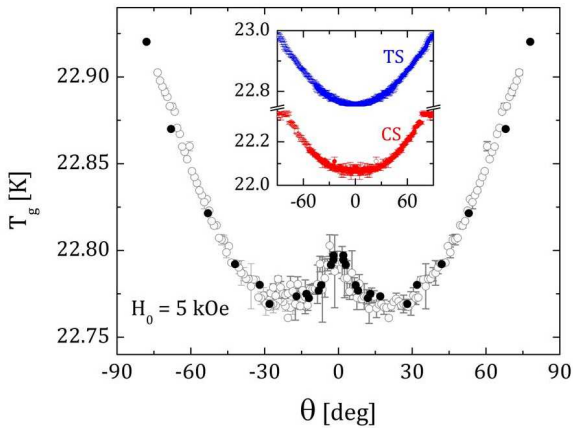}
\caption{\footnotesize{(Color online) Vortex liquid-to-glass transition
temperature $T_g$ as a function of the orientation $\theta$ of an applied
field of $5\ \text{kOe}$  with respect to the ${c}$ axis for a free sample (F , main panel) and for strained and compressed  strain conditions (TS and CS, inset). Glass temperatures were obtained by means of a nonlinear fit
according to the reported model \cite{Ghorbani2012} of the temperature
dependence of $\rho$. Black solid circles stand for $T_g$ obtained from
curves $\rho(T)$ measured in temperature ramps at fixed $\theta$ with temperature step of $\Delta T = 1\ \text{mK}$,  and gray open circles correspond to measurements obtained in a single cooling temperature ramp while periodically rotating the field direction. Error bars represent the $95\%$ confidence
intervals for the estimated $T_g$.} }
\label{fig:Tglass}
\end{figure}

The liquid-glass temperature transition $T_{g}$ can be obtained by means of a nonlinear fit, taking into account the  critical behavior of the resistivity $\rho(H,T)$ at $T_{g}$. An alternative way \cite{Ghorbani2012} proposes a scaling of the resistivity $\rho(H,T)$ by assuming that the glass transition occurs when thermal and pinning energy scales match. This scaling procedure has been performed at $H_{0}=5\ \text{kOe}$, in order to obtain the best $T_{g}$ for each $\theta $ direction, and under different strain conditions (F, CS and TS). The main panel in Fig. \ref{fig:Tglass} presents the angular dependence of $T_{g}$ in the F condition. The local maximum of $T_{g}$ for angles close to the ${c}$ axis
violates the expected angular dependence in the presence of uncorrelated
random disorder. A maximum in $T_{g}(\theta )$ at $\theta =0$ was also observed in twinned cuprate superconductors and associated with a transition to a Bose glass phase \cite{Nelson1992,Grigera98}. Conversely, the anomaly is absent in the TS and CS conditions (see the inset),
consistent with a single domain detwinned sample.

With the aim being to quantify the underlying anisotropy and to be able to compare results obtained in different conditions, we make use of the \textit{effective
field} concept. Essentially, the idea is to find the magnetic field that would be necessary to apply, in a hypothetical isotropic situation, in order to obtain the measured resistivity in the real anisotropic case. In materials with uniaxial anisotropy the dependence of any
property on the direction of the magnetic field relative to the ${c}$ axis $Q(H_{0},\theta )$ can be related to the magnetic field dependence $Q(H,\theta _{0})$ through an \textit{effective field} $H_{\text{eff}}(\theta )$
defined as $Q(H_{0},\theta )=Q(H_{\text{eff}}(\theta ),\theta _{0})$. $H_{\text{eff}}$ is
well defined if $Q(H)$ is a one-to-one function. In a variety of
experiments, carried out with different techniques in tetragonal (or
slightly orthorhombic twinned) type-II superconducting materials \cite{Herbsommer2000, Marziali2013, Jaros2008}, the
angular dependence of $H_{\text{eff}}$ is well fitted by
\begin{equation}
\frac{H_{\text{eff}}}{H_{0}}=\frac{\sqrt{\gamma ^{2}\cos^{2}\theta +\sin^{2}\theta }}{%
\sqrt{\gamma ^{2}\cos^{2}\theta _{0}+\sin^{2}\theta _{0}}}.
\label{eq:campo efectivo}
\end{equation}
where the constant $\gamma$ characterizes the uniaxial anisotropy. This
dependence holds if randomly distributed defects are the prevailing source
of vortex pinning, but breaks down when the predominant pinning is due to
correlated defects such as {DBs}, $ab$ planes, or columnar defects. However, in a single orthorhombic domain,
different $\gamma _{a}$ and $\gamma _{b}$ constants could, in principle,
hold.

\begin{figure}
\centering
\includegraphics[width=\linewidth]{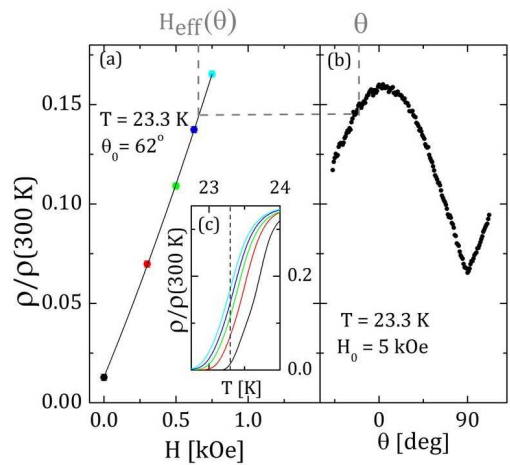} 
\caption{
\footnotesize{ (Color online) Illustration of the effective field concept: (a) field-dependent normalized resistivity $\rho/ \rho(300\ \text{K})$ for $\theta = 62^{\circ}$ and $T = 23.3\ \text{K}$ in TS conditions. Data points were obtained, as indicated in (c) by the intersection with the dotted vertical line, from the temperature dependent curves for different magnetic fields. Each color [symbol in (a) and curve (c)] corresponds to a different magnetic field between $0$ and $7.5\ \text{kOe}$. (b) Normalized resistivity as a function of the orientation of $H =5\ \text{kOe}$ for the same temperature as in (a).
} }
\label{fig:Metodo_Mariano}
\end{figure}

To obtain the effective field from our results, we have
complemented the data shown in Fig. \ref{fig:Polares4} with measurements
of the magnetoresistivity as a function of the intensity of the magnetic
field. As plotted in the inset of Fig. \ref{fig:Metodo_Mariano}, the
sample was field cooled under different constant magnetic fields ranging
between $0$ and $7.5\ \text{kOe}$, applied at a fixed angle, away from the DBs. Figure \ref{fig:Metodo_Mariano}(a) shows $\rho (H)$ for $\theta _{0}=62^{\circ }
$ at $T=23.3\ \text{K}$; Fig. \ref{fig:Metodo_Mariano}(b) shows $\rho (\theta )$ at $%
H_{0}=5~\text{kOe}$ at the same temperature and the same strain conditions (TS in this example). We then obtain $H_{\text{eff}}$ in the same way as in Ref. \citenum{Marziali2013}, identifying the intensity $H$ needed in Fig. \ref{fig:Metodo_Mariano}(a) to
match the resistivity value at a given angle $\theta $ in Fig. \ref{fig:Metodo_Mariano}(b) (see
gray labels on top axis), $\rho \left( H_{\text{eff}},\theta _{0}\right) =\rho \left(
H_{0},\theta \right) $.

\begin{figure}[h!]
\centering
\includegraphics[width=\linewidth]{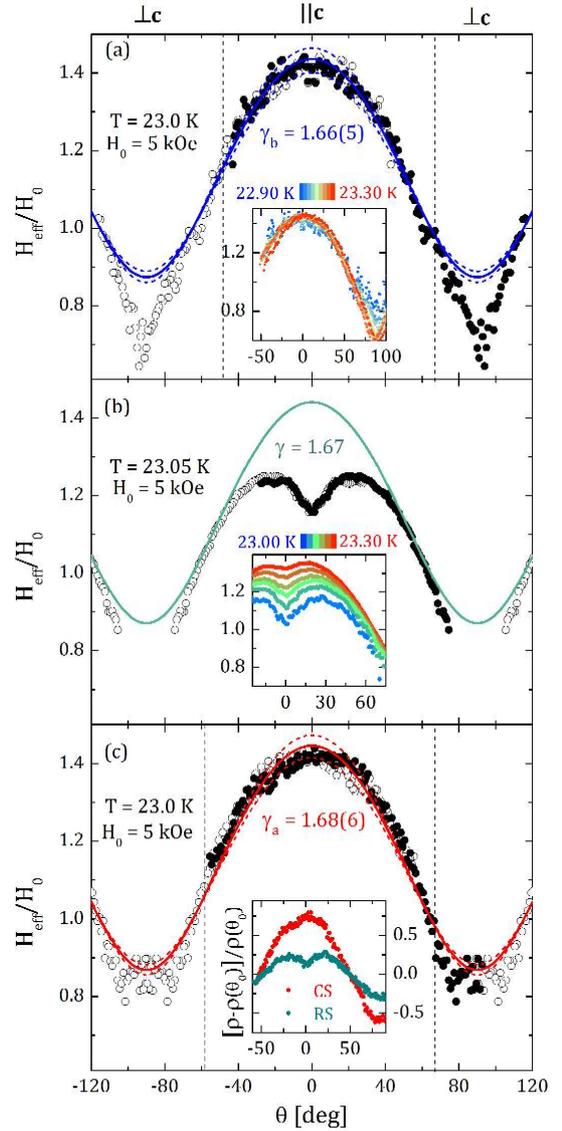} 
\caption{\footnotesize{(Color online) $H_{\text{eff}}$ as a function of applied field orientation $\theta$, for the same sample under (a) tensile, (b) null, and (c) compressive strain, at $T \approx 23.0\ \text{K}$. $H_{\text{eff}}$ is normalized by the applied field $H_0=5\ \text{kOe}$. Measured data are represented in solid circles, while open symbols stand for equivalent data points generated by symmetry for completion. Best-fitting curves are plotted by blue and red solid lines according to Eq. (\ref{eq:campo efectivo}), in (a) and (c), respectively, and colored dashed lines account for uncertainty in the estimated parameter ($95\%$ confidence intervals). Data corresponding to field directions near the $ab$ planes beyond the vertical dashed lines were excluded from the fit. The observed angular dependence for F in (b) is contrasted to the one predicted by Eq. (\ref{eq:campo efectivo}) with $\gamma=1.67$, shown by the green solid line. The angular dependence of $H_{\text{eff}}$ for temperatures ranging from $22.90$ to $23.30\ \text{K}$ is included in the insets of (a) and (b) for TS and F, respectively. The inset of (c) presents the angular dependence of the normalized resistivity $[\rho-\rho(\theta_0)]/\rho(\theta_0)$ of the sample after the strain was released (green symbols) compared to that obtained for CS (red symbols), both at $T=23\ \text{K}$ and $\theta_0=62^{\circ}$.} } 
\label{fig:Scaling}
\end{figure}

Figure \ref{fig:Scaling} presents the angular dependence of $H_{\text{eff}}$ for
different strain conditions for the same sample. Good agreement with the
scaling function in Eq. (\ref{eq:campo efectivo}) holds, as shown by blue and red solid lines,
over a wide angular range for TS [Fig. \ref{fig:Scaling}(a)] and CS [Fig. \ref{fig:Scaling}(c)], but scaling definitely
fails to reproduce the observed behavior in F [Fig. \ref{fig:Scaling}(b)], especially when the
orientation of the field is close to the ${c}$ axis. This anomaly is
reinforced as the temperature is lowered, as depicted in the inset in Fig. \ref{fig:Scaling}(b), while the temperature dependence in the strained sample conditions, TS
and CS, is negligible [see the inset in Fig. \ref{fig:Scaling}(a)]. The absence of the dip near small angles in the TS and CS conditions, together with the good fit with the scaling function in Eq. (\ref{eq:campo efectivo}), further supports that the sample was successfully detwinned with the applied strains $(0.26\pm 0.05)\ \%$ and $(-0.35\pm 0.05)\ \%$. In addition, we observed that the signature of DBs reappeared as the strain was released in the orthorhombic nematic phase \cite{Tanatar2010}, as shown in the inset in Fig. \ref{fig:Scaling}(c) for $[\rho (\theta)-\rho (\theta _{0})]/\rho (\theta _{0})$.

Under the premise of achieved detwinning, the shortest lattice constant $b$
would align with the compressed direction, while the lattice constant $a$
would align with the elongated direction. In that sense, the measured
resistivity would correspond to $\rho _{b}$ in the case of CS, and to $\rho
_{a}$ in the case of TS. It should be kept in mind that these resistivities are
additionally affected by the corresponding tensile (compressive) strain,
and the corresponding compression (expansion) in the transverse directions. The inset of Fig. \ref{fig:Trancisiones_strain} presents the resistivity anisotropy obtained from both magnitudes, larger than $0.4$ throughout the whole temperature range in which the resistivity can be measured. 
The coincident resistivities obtained for free conditions (F and SF), on the other hand,
correspond to an average of $\rho _{a}$ and $\rho _{b}$ under zero strain
\cite{Tanatar2016} in the temperature range where resistivity is unaffected by DBs.

Along with the line of reasoning that we have achieved a single oriented
domain in TS and CS conditions, we conclude that the magnetic field is
rotated in the $cb$ and $ca$ planes in each strain condition. Therefore, the
best-fit parameters shown in Fig. \ref{fig:Scaling} are $\gamma _{b}$ and $%
\gamma _{a}$, respectively, with no discernible differences ($\gamma
_{b}=1.66\pm 0.05$, $\gamma _{a}=1.68\pm 0.06$) \cite{comentario_razonable}.

Blatter \textit{et al.} \cite{Blatter} explained the angular
dependence of the effective field in superconductors with {\textit{uniaxial}}
anisotropy, in the context of a Ginzburg-Landau (GL) free-energy functional \cite{GL, Tinkham}
by proposing simple rules that scale the anisotropic problem into an isotropic
situation. In that work, anisotropy is modeled by 
introducing coefficients in the gradient terms of the free-energy expansion as
\begin{equation}
\mathcal{G}_{\text{sc}}=\int {\text{d}^{3}\mathbf{r} \sum_{\mu}{\frac{1}{2 m_{\mu}} \vert D_\mu \Delta \vert ^2}+ \alpha \vert \Delta \vert ^2+
\beta \vert \Delta \vert ^4}
\end{equation}
where $\Delta$ is the complex order parameter, $\alpha$ and $\beta$ are the standard parameters in the free-energy expansion, and
 $D_\mu=\frac{\partial}{\partial x_\mu}+\frac{ie^*}{\hbar c} A_\mu$ is the covariant derivative. The index $\mu$ runs over the crystal axes $a,b,c$.
 The parameters $m_\mu$ are related
to the {\em phase stiffness} \cite{Benfatto2001, Emery1995} via $\hbar^2|\Delta_0|^2/m_{\mu}$, where $|\Delta_0|$ stands for the $T=0$ bulk value of the order parameter.
Given that orthorhombic distortion in these materials is ascribed to nematic ordering, it is reasonable to assume that $m_a = m_b = m_{\vert \vert}$ in the absence of nematicity. In that case, Blatter \textit{et al.} have shown that the parameter $\gamma$ of Eq. (\ref{eq:campo efectivo}) can be identified
with $\gamma=\frac{m_c}{m_{\vert \vert}}$.

When a nematic order parameter $\eta$ is present, we can expect the usual bi-quadratic coupling
\begin{equation}
\mathcal{G}_{\text{int}} =\lambda_2 \int {{d}^{3}\mathbf{r}\ \eta^2 \vert \Delta\vert^2}
\end{equation}
which will directly affect the value of $T_c$. In addition, we expect the presence of a term coupling the nematic order to the gradient of the superconducting order parameter of the form

\begin{equation}
\mathcal{G}_{\text{int}}^1=\frac{\lambda_1}{2} \int {{d}^{3}\mathbf{r}\  \eta \left\{ \left\vert D_a \Delta \right\vert ^{2}-\left\vert D_b \Delta \right\vert ^{2}\right\} }.
\end{equation}
In the case of a detwinned sample with $\eta=\eta_0$ this amounts to working with a standard GL model except
with 
\begin{equation}
\frac{1}{m^*_{a,b}}= \frac{1}{m_{\vert \vert}} \pm \lambda_1 \eta_0.
\label{eq:modificacion_m}
\end{equation}

This fact immediately suggests the generalization of Eq. (\ref{eq:campo efectivo}), but now with two \textit{a priori} different $\gamma_a$ and $\gamma_b$ depending on whether the external field is rotated in the $ac$ or $ab$ plane.
Our results show that there is no significant difference in the values of $\gamma_a$ and $\gamma_b$, implying that although nematicity contributes in a decisive way to the conduction properties, it does not strongly affect the in-plane anisotropy of the stiffness parameters.
From Eq. (\ref{eq:modificacion_m}), the latter could be related either (a) to a suppressed $\eta_0$ due to the proposed competition with superconductivity \cite{Nandi2010, Kalisky2010} or (b) to a weak coupling constant $\lambda_1$. For item (b) to be true, nematic superconducting coupling should then be expressed in higher-order even coupling terms such as $\lambda_2 \eta_0^2|\Delta|^2$.

In order to predict the angular dependence of superconducting properties, Blatter and coworkers \cite{Blatter, BlatterB} further proposed additional scaling rules. In particular, the resistivity is predicted to scale in proportion to the parameter $m_{\mu }$. 
In the present case, however, ${m}_{a}\simeq {m}_{b}$ in spite of the strong anisotropy between the resistivities $\rho _{a}$ and $\rho _{b}$.

 The observed mismatch between the in-plane anisotropies of superconducting stiffness and resistivity indicates that the symmetry of the transport properties in the mixed phase is strongly influenced by the normal-state in-plane anisotropy. Moreover, it is reminiscent of the mismatch between the out-of-plane anisotropies of normal effective mass and resistivity measured in various iron superconductors \cite{GLTD}. The latter suggests that sources of anisotropy other than the Drude weight \cite{Mirri2015}, such as anisotropic scattering \cite{Allan2013, Tanatar2016}, could be relevant in the normal and superconducting transport properties. 
 
 The strong difference in dissipation between TS and CS samples in the mixed superconducting vortex liquid phase, is qualitatively consistent with both the observed difference in $T_g$ and the reported critical current in-plane anisotropy in the absence of DBs for similar compositions \cite{Hetcher2018}. Hecher \textit{et al.} attribute the observed $J_{c}$ anisotropy either to an anisotropic pinning efficiency induced by Co impurities or to the multiband electronic structure of these compounds. An insight favoring this last possibility is the subtle loss of the $C_2$ symmetry observed at the onset of the resistive transition for some strain conditions. Careful observation of the color maps in the polar plots in Fig. \ref{fig:Polares4}, shows that there is a weak loss of  $C_2$  symmetry, similar to that recently observed in in-plane angular-dependent resistivity and associated with the multiband character \cite{Li2018}. This measured effect (within the circular segments delimited by red lines) is more evident  at higher temperatures, i.e., at the largest radius in the polar plots, where the angular dependence of the resistivity is very small (light blue).

 Full time-dependent GL calculations, with the appropriate Legendre transform (taking into account the fact that in transport experiments the current is one of the independent variables \cite{Tinkham}) are required for an accurate description.

\section{Conclusions}
\label{sec:Conclusions}

The anisotropy in the superconducting transport properties was
investigated in slightly underdoped $\text{Ba}(\text{Fe}_{1-x}\text{Co}_{x})_{2}\text{As}_{2}$ single
crystals by measuring the resistive superconducting transition under the
rotation of an applied magnetic field and different strain conditions. We
were able to reproduce the free sample response under controlled uniaxial
stress and, consequently, measure the absolute sample deformation. We detected
the presence of DBs in free samples from the breaking of the expected intrinsic angular dependence for a single orthorhombic domain: an anomaly was detected in the vortex liquid-glass transition temperature $T_{g}$, as well
as in the angular dependence of the resistivity in the vortex liquid phase.
The suppression of this anomaly, indicative of the sample detwinning, was
achieved under the application of strong compressive and tensile strains $%
\varepsilon \lesssim $ $0.3\%$, considerably higher than the reported orthorhombic distortion. 

{For the samples employed, there is no magnetic long-range order; thus, only nematic and superconducting order parameters need to be taken into account.} By extending a Ginzburg-Landau scaling formalism to nematic systems and coupling at first order the nematic and superconducting order parameters, we obtained the in-plane superconducting stiffness anisotropy under strain. 
Our results show no significant differences between the superconducting stiffness in the orthorhombic $a$ and $b$ axes. 
On the other hand, a strong in-plane resistivity anisotropy holds in the mixed superconducting phase, indicating that normal and/or nonequilibrium properties are playing a key role.
Under the GL formalism, the lack of a measurable in-plane stiffness anisotropy coexisting with a clear in-plane resistivity anisotropy is surprising and requires further investigation.

\ An additional important remark is that part of these conclusions is based on the assumption of the validity of the GL formalism, under debate in these materials \cite{Sachev}. Assuming this framework, full time-dependent GL calculations, as well as a model for the interaction between nematic DBs and superconducting vortices, are necessary to rigorously quantify our experimental results.

\textbf{Acknowledgements}: We would like to acknowledge Ni Ni for initial growth of these samples. The authors are specially grateful to Santiago Grigera, Rodolfo Borzi and Andrew Mackenzie for providing the strain apparatus used in the reported experiments and for technical support  
as well as for useful discussions. We also thank Mark Barber for providing us accessories and useful tips, and Maricel Rodríguez and Florencia Di Salvo for facilitating the X-ray characterization. This work was partially supported by CONICET and Universidad de Buenos Aires and  the US National Science Foundation through grant No. DMR 1725401 at the University of Illinois (EF).  Work done at Ames Laboratory (PCC) was supported by the U.S. Department of Energy, Office of Basic Energy Science, Division of Materials Sciences and Engineering.  Ames Laboratory is operated for the U.S. Department of Energy by Iowa State University under Contract No. DE-AC02-07CH11358. EF thanks the Department of Physics of the Universidad de Buenos Aires for its hospitality.

\end{document}